# Record High $T_c$ Element Superconductivity Achieved In Titanium

Changling Zhang[a,1,2], Xin He [a,1,2,3], Chang Liu[a,4], Zhiwen Li[1,2], Ke Lu[1,2], Sijia Zhang[1], Shaomin Feng[1], Xiancheng Wang*[1,2], Yi Peng[1,2], Youwen Long[1,2,3], Richeng Yu[1,2], Luhong Wang[5], Vitali Prakapenka[6], Stella Chariton[6], Quan Li[4], Haozhe Liu[7], Changfeng Chen*[8], Changqing Jin*[1,2,3]

[1]Beijing National Laboratory for Condensed Matter Physics, Institute of Physics, Chinese Academy of Sciences, Beijing 100190, China
[2]School of Physical Sciences, University of Chinese Academy of Sciences, Beijing 100190, China
[3] Songshan Lake Materials Laboratory, Dongguan 523808, China
[4]International Center for Computational Method and Software, College of Physics, Jilin University, Changchun 130012, China
[5] Shanghai Advanced Research in Physical Sciences, Shanghai 201203, China
[6] Center for Advanced Radiations Sources, University of Chicago, Chicago, Illinois 60637, USA
[7] Center for High Pressure Science & Technology Advanced Research, Beijing 100094, China
[8] Department of Physics and Astronomy, University of Nevada, Las Vegas, Nevada 89154, USA

## Abstract

It is challenging to search for high $T_c$ superconductivity (SC) in transition metal elements wherein $d$ electrons are usually not favored by conventional BCS theory. Here we report discovery of surprising SC up to 310 GPa with $T_c$ above 20 K in wide pressure range from 108 GPa to 240 GPa in titanium. The maximum $T_c^{onset}$ above 26 K and zero resistance $T_c^{zero}$ of 21 K are record high values hitherto achieved among element superconductors. The $H_{c2}(0)$ is estimated to be ~ 32 Tesla with coherence length 32 Å. The results show strong $s$-$d$ transfer and $d$-band dominance, indicating correlation driven contributions to high $T_c$ SC in dense titanium. This finding is in sharp contrast to the theoretical predications based on pristine electron-phonon coupling scenario. The study opens a fresh promising avenue for rational design and discovery of high $T_c$ superconductors among simple materials via pressure tuned unconventional mechanism.

[a] These authors contributed equally to the paper
* Corresponding authors; wangxiancheng@iphy.ac.cn; changfeng.chen@unlv.edu; jin@iphy.ac.cn

# Introduction

Titanium (Ti) metal has long attracted tremendous scientific interests because of its combined properties of light weight, high strength and corrosion resistance. As an advanced metallic structural material, Ti and its alloys find wide use in the fields of aerospace, biomedicine and at extreme conditions [1-3]. High pressure can modify crystal structures which, in turn, may lead to new functionalities. At ambient pressure and room temperature, Ti crystalizes in a hexagonal close-packed (hcp) structure (Ti-α phase) [4]. Under applied pressure, Ti undergoes structural transitions in the sequence of Ti-α, Ti-ω, Ti-γ, Ti-δ and Ti-β phases, where Ti-ω phase is a hexagonal structure, Ti-γ and Ti-δ phases are orthorhombic and Ti-β phase is body-centered cubic [5-9]. The α-to-ω transition occurs around 8 GPa [5, 6], and the Ti-ω phase is stable up to about 100 GPa, then transforms into Ti-γ phase [6, 10], which further transforms into Ti-δ phase at ~140 GPa [6], before cubic Ti-β phase stabilizes at 243 GPa [9].

Superconductivity (SC) in high-pressure phases of Ti metal was previously reported to have a measured maximal critical temperature ($T_c$) of 3.5 K at 56 GPa [11]; early theoretical calculations based on the electron-phonon coupling mechanism predicate that the $T_c$ for Ti metal is capped at about 5 K for all the known high pressure phases [12]. Generally, compression of crystal lattice has markedly different effects on the 4$s$ and 3$d$ bands, prompting notable $s$-$d$ electron transfer. The narrower $d$-bands possess stronger correlation characters, while the $s$-$d$ transfer tends to enhance electronic density of state (DOS) near the Fermi level in favor of SC [13-17]. Here, we report a surprising experimental observation of dramatic pressure enhanced SC in Ti over a wide pressure range with the maximal $T_c$ above 26 K, setting a new record among elemental superconductors. Measured normal-state conductivity results and analysis of electronic and superconducting properties indicate prominent influence of the *s-d* transfer and *d*-band driven correlation effects on the significantly enhanced and unusually high $T_c$ SC in highly compressed Ti metal.

## Results

**High pressure superconductivity under high pressure.** The high pressure resistance experiments have been carried out using diamond anvil cell apparatus. The sample assembly and the electrode configuration for four-probe Van der Pauw method are shown in Supplementary Figure 1a-b. The pressure is calibrated by the shift of the first order Raman edge frequency from the diamond cutlet as shown in Supplementary Figure 1c. Fig. 1a shows the experimentally measured temperature dependence of electrical resistance under high pressure up to 180 GPa for Ti metal in Sample 1. The results show that Ti metal becomes superconductive with $T_c$ above 2 K at 18 GPa and rises slightly to ~3.5 K at 54 GPa, which agrees with previously reported results[11]. The value of $T_c$ increases at an enhanced pace from 54 to 99 GPa, then undergoes a steep rise from 10.2 K at 99 GPa to 20.3 K at 108 GPa, and $T_c$ is further enhanced to 22 K at 180 GPa. For Sample 2, the pressure range was increased, with the highest pressure reaching 310 GPa that was calibrated by using the method described in Ref. 18. The resistance curves under pressure are shown in Fig. 1b. With further increasing pressure the $T_c$ reaches a maximum of 26 K at 248 GPa, as shown in Fig. 1c. The onset critical temperature ($T_c^{onset}$), the midpoint critical temperature ($T_c^{mid}$) and the zero-resistance critical temperature ($T_c^{zero}$) of superconducting transition are determined by the derivative of resistance with respect to temperature $dR/dT$ as shown in the inset of Fig. 1c. It is noted that $T_c$ stays almost constant over a wide pressure range up to at least 240 GPa, and such robust superconducting behaviors are consistently seen in different specimens (Supplementary Figure 2). At 108 GPa where $T_c$ jumps up, the resistance exhibits a two-step superconducting transition behavior with the lower $T_c$ of ~11 K that is comparable to the $T_c$ value at 99 GPa, indicating a phase transition near 108 GPa, which is close to the pressure for the reported ω-γ phase transition [6, 10]. In the Ti-ω phase, $T_c$ rises smoothly with pressure below 56 GPa with a slope of 0.07 K/GPa; assuming this slope keeps unchanged, $T_c$ should reach 8.7 K before the ω-γ phase transition (at 128 GPa) [11]. However, our results show that the slope $dT_c/dP$ increases significantly between 54-99 GPa, which leads to the much higher measured $T_c$ at 99 GPa.

To further probe the pressure driven SC phase of Ti metal, we have examined the effect of magnetic field on the SC transition behavior. Fig. 2a presents the electrical resistance measured at 248 GPa and under different magnetic fields. It is seen that the transition is gradually suppressed by the magnetic field. We have plotted onset $T_c$ versus magnetic field in Fig. 2b, from which the upper critical field at zero temperature $\mu_0H_{c2}(0)$ can be estimated. The $\mu_0H_{c2}(T)$ date were fitted to the Ginzburg Landau (GL) function,

$$\mu_0H_{c2}(T) = \mu_0H_{c2}(0)(1 - (T/T_c)^2) \quad (1)$$

which gives a value of $\mu_0H_{c2}(0)$=32 T. At other pressures where $T_c$ is above 20 K, the estimated upper critical field has been obtained to be near 30 T (Supplementary Figure 3a-b), which is larger than that of the most commonly used low-temperature NbTi superconductor ($\mu_0H_{c2}(T)$=15 T). Using the $\mu_0H_{c2}(T)$ value of 32 T, the GL coherence length was calculated to be ξ=32 Å via $\mu_0H_{c2}(0)= \Phi_0/2\pi\xi^2$, where $\Phi_0$= 2.067×10$^{-15}$ Web is the magnetic flux quantum.

**Superconducting phase diagram**. Fig. 3a and b present the measured Hall resistance at room temperature and different pressures. The Hall resistance is negative and decreases linearly with magnetic field, indicating that the dominant carriers are electrons. The carrier density $n$ as a function pressure estimated from the Hall resistance is presented in Fig. 3c, with a value of about $2.5 \times 10^{22}/cm^3$ that is little changed in the low-pressure range. Above 108 GPa, $n$ increases dramatically and is enhanced by more than one order of magnitude to $3.1 \times 10^{23}/cm^3$ at 137 GPa. Further increasing pressure leads to reduced carrier density of $4.5 \times 10^{22}/cm^3$ at 144 GPa. The changes of the carrier density indicate phase transitions at pressures of about 108 GPa and 144 GPa, respectively. To see more clearly these phase transitions, the pressure dependence of resistance $R(P)$ at fixed temperature is plotted in Fig. 3d. The $R(P)$ curve shows two peaks near the critical pressures, which corresponds to the ω-γ phase and γ-δ phase transitions reported by Y. Akahama [6] and Y. K. Vohra [10], respectively.

We have carried out the high-pressure X-ray diffraction experiments, and the results are shown in Supplementary Figure 4a-d. Combining the phase transition reported by previous works and our transport experiments, we plot the superconducting versus structural phase diagram in Fig. 4. Up to 310 GPa, five different crystal structures, in the sequence of Ti-α, Ti-ω, Ti-γ, Ti-δ and Ti-β are identified. The Ti-α phase ($P$=0~9 GPa) hosts SC with $T_c$ below 2 K; the Ti-ω phase ($P$=9~108 GPa) sees a monotonously increasing $T_c$ with pressure to 12 K at ~108 GPa; while in the high-pressure phases of Ti-γ ($P$=108~144 GPa) and Ti-δ ($P$=144~240 GPa), $T_c$ stays above 20 K with the maximum $T_c$ =26 K occurring at the boundary of Ti-δ and Ti-β phases. It is remarkable that the $T_c$ of compressed Ti metal stays robust above 20 K in a very wide range of pressures. This behavior is at odds with the expectation of conventional phonon-mediated superconducting theory, which predicts sensitive pressure dependence and descending $T_c$ at very high pressures due to phonon stiffening. Interestingly, it is noted that the phenomenon of robust $T_c$ values extended over a wide pressure range has been found in high-$T_c$ cuprates wherein SC is driven by correlation effects [19], and this analogy indicates that mechanisms of novel many-body origin may be driving the unexpectedly SC in dense Ti metal at ultrahigh pressures.

## Discussion

Among all elemental solids, only a few have been reported to exhibit SC with $T_c$ near 20 K at high pressures, including alkali metal Li [20, 21], alkali-earth metal Ca [15] and rare-earth metals of Y [16]. The high $T_c$ values of Li, which is the lightest elemental metal containing only simple 2$s$-electrons, is attributed to the enhancement of the electron-phonon coupling due to the phonon modes softening under high pressure [22]; while the $T_c$ enhancements in Ca and Y are mainly ascribed to pressure induced electron transfer from the $s$ to $d$ shell. The superconducting Ti under high pressure has the highest $T_c$ among the elements in the period table.

The resistance of Ti as a function of temperature measured at 18 GPa (Supplementary Figure 5) exhibits interesting scaling behavior. Upon fitting with the formula $R = R_0 + A*T^n$, where $R_0$ is the residual resistance, $A$ is the coefficient of the power law, and $n$ is the exponent, the resulting scaling exponent $n$=3.1 is different from the value expected for systems dominated by either electron-phonon scattering ($n$=5) or pure electron-electron scattering ($n$=2). In fact, the $n$ value near 3 implies that the $s$-$d$ interband scattering dominates the electron transport with contributions from electron correlation effects, as observed in 1T-TiSe$_2$ [23] or Ta$_4$Pd$_3$Te$_{16}$ [24]. It reveals that at 18 GPa the energies between 4$s$ and 3$d$ shells is very near to each other and their electron configurations are mixed. Upon further applying pressure, the $s$ band will move to higher energy as shown in the following calculations, promoting the $s$-$d$ electron transfer, increasing the number of $d$ electrons and inducing sequence of structural phase transitions. As a result, the $T_c$ is greatly enhanced and a record $T_c$ among the elements has been achieved. It seems that the $T_c$ is closely related with the $s$-$d$ electron transfer, which suggests the impact of the $d$ electron correlation effects on the formation of Cooper pairs.

We have examined relative energetic stability of the high-pressure phases of Ti metal from first-principles calculations, and the results are consistent with the experimentally measured phase stability and transformation sequence. Adopting the determined crystal structures of the Ti metal phases, we have calculated their electronic, phonon and electron-phonon coupling properties, which are used as input to determine the SC critical temperature $T_c$. The obtained $T_c$ data for the Ti-ω, Ti-γ, and Ti-δ phases (Supplementary Figure 6) reveal the following trends: (i) $T_c$ increases with rising pressure monotonically in the Ti-ω phase over its entire stability range; (ii) $T_c$ is significantly enhanced upon the phase transition into the Ti-γ phase over the relatively narrow pressure range where this transition occurs; (iii) $T_c$ undergoes an even larger jump when the structure enters the Ti-δ phase. These findings are in overall agreement with the experimental results up to about 160 GPa, but large discrepancies exist at higher pressures where the record-setting $T_c$ is observed.

The experimentally observed superconducting properties of the densely compressed Ti metal indicate clear inadequacy of the conventional phonon-mediated

SC mechanism for describing the unexpectedly high values of $T_c$ and anomalously robust superconducting state over a very wide pressure range. The electron-electron correlation effects associated with the *d*-bands in Ti are expected to have a strong influence on transport and superconducting properties. Calculated electronic band structure of Ti-ω phase at 20 GPa (Fig. 5a) shows significant overlap of the *s*- and *d*-electron states near the Fermi level, which corroborates our measured normal-state resistivity results (Supplementary Fig. 5) indicating strong *s-d* scattering contribution to the resistivity [24, 25]. At higher pressure of 100 GPa, the band structure of Ti-γ phase (Fig. 5b) shows that the *s*-states move up in energy while still have large overlap with the *d*-bands, which is conducive to significant *s-d* scattering. At further increased pressure of 180 GPa, the electronic states in Ti-δ phase (Fig. 5c) near the Fermi level are dominated by the *d*-electron states. There are flat *d*-bands below the Fermi level that can host significant correlation effects. These *d*-bands will likely rise much closer to the Fermi level due to the electron-electron interactions when properly treated by many-body theory. Such correlation effects could greatly enhance the *d*-band derived electronic DOS at the Fermi level favorable for increasing $T_c$ and also drive additional non-phonon-mediated mechanisms for further strengthening the superconducting state with higher $T_c$ and maintaining its robust presence over a wide pressure range. This study raises intriguing possibility of major impact by many-body effects on SC in dense Ti, which needs in-depth study by sophisticated many-body theoretical treatments to explore pertinent novel processes and underlying mechanisms.

The present study unveils unexpectedly high $T_c$ SC in compressed Ti metal with record-setting among elemental superconductors. The $T_c$ and $\mu_0 H_{c2}$ values of Ti are notably higher than those of the widely used superconducting NbTi alloy. Our discovery raises the possibility of finding more materials via pressure driven correlation effects stemming from the contributions of *d* electrons, leading to SC with much higher $T_c$ than previously believed achievable, and such materials may be stabilized at lower pressures via mechanical strain or chemical pressure. This intriguing scenario calls for further research into the impact on SC by the *s-d* interaction and *d*-electron dominated correlation effects in highly compressed *d*-band metals for evaluation of diverse materials, from elemental solids to alloys, in search of hitherto unknown and unexplored superconducting materials that could improve fundamental understanding of broader varieties of superconductors. Equally important, the present findings open new avenues for expanding the scope of superconductors with notably enhanced $T_c$ and $\mu_0 H_{c2}$ that are more adaptive and suitable for applications in diverse and demanding implementation settings. We became aware during preparing the paper that an independent work by Liu X.Q. *et al*. was carried with the similar results[25].

## Methods

**High pressure measurements.** The electrical resistance and Hall Effect measurements were performed using the four-probe Van der Pauw method as shown in Supplementary Figure 1b. The pressures are calibrated via the shift of the first order Raman edge frequency from the diamond cutlet as shown in Supplementary Figure 1c[18]. The applied current is 100 μA. Diamond anvil cells were used to produce high pressures. A variant of anvils with beveled culet size of 20/140/300μm, 30/140/300μm or 50/140 300μm are adopted in the experiments. A plate of T301 stainless steel that is covered with mixture of cBN powder & epoxy as insulate layer was used as the gasket. A hole of approximately 15 to 30 μm in diameter depending on top culet size was drilled in the center of the gasket to serve as high pressure chamber. The hBN powder was generally used as pressure-transmitting medium that filled in the high-pressure chamber. We used the ATHENA procedure to produce the specimen assembly[26]. Four Pt foils with thickness approximately 0.5 μm as the inner electrode were deposited on the culet surface. Cross-shaped Ti specimens with side lengths ~ 10μm*10μm and thickness of 1 μm were stacked on the electrodes. Pressure was calibrated by the shift of the first-order Raman edge frequency from the diamond cutlet[18]. Diamond anvil cells were put in a MagLab system that provides synergetic extreme environments with temperatures from 300 K to 1.5 K and magnetic fields up to 9 T for the transport measurements [26-29].

**High-pressure synchrotron X-ray experiments**. *In-situ* high-pressure angle-dispersive X-ray diffraction data were collected at room temperature at GSECARS of Advanced Photon Source at the Argonne National Laboratory. The x-ray with the wavelength λ = 0.3344 Å was focused down to a spot of ~3 μm in diameter on the sample. A symmetric diamond anvil cell with beveled anvil (50/300 μm) was used. Rhenium steel gasket was pre-pressed to a thickness of 20 μm, and a hole of diameter of 15 μm was drilled at the center to serve as sample chamber, which was then filled with Ti power mixed with Pt. Pressure was calibrated using the equitation of state of both Re and Pt. The X-ray diffraction images are converted to two dimensional diffraction data with Dioptas[30].

**First-principle theoretical calculation.** To assess structural, electronic, and phonon-mediated superconducting properties of Ti metal under pressure, we employed the latest computational techniques to determine the total energy, lattice dynamics and electron-phonon coupling using the QUANTUM ESPRESSO code[31], with improved description over previously reported results[12]. Superconducting critical temperature $T_c$ has been evaluated based on the Eliashberg theory of SC [32, 33], using the following formula that McMillan derived[34] and later modified by Allen and Dynes [35],

$$T_c = \frac{\omega_{log}}{1.20} exp[-\frac{1.04(1+\lambda)}{\lambda - \mu^*(1+0.62\lambda)}]$$

where $\omega_{log}$ is a logarithmically averaged characteristic phonon frequency, and $\mu^*$ is the Coulomb pseudopotential which describes the effective electron-electron repulsion [36]. This equation is generally accurate for materials with EPC parameter $\lambda$ at 1.5 or less [37,38], which is satisfied in the present study. The Coulomb pseudopotential $\mu^*$ is often treated as an adjustable parameter with values within a narrow range around 0.1 for most materials, making this formulism highly robust [36-39], and compares well with the latest *ab initio* Eliashberg theory [40]. In this work, the commonly used value of $\mu^*$=0.13 is adopted for all the reported calculations. Such pseudopotential calculations have been employed to study structural stability and transformation of Ti compressed up to at least 200 GPa, and the results are in good agreement with those from full potential calculations and provide a good description of the experimental data[4]. We also calculated electronic band structures to assess the evolution of the *s*- and *d*-bands under pressure.

## Figure Captions

**Fig. 1. The superconductivity measurements. a** Temperature dependence of the electrical resistance of elemental metallic Ti (Sample 1) measured under high pressures. The inset is an enlarged view of the resistance curve, showing the superconducting transition in detail. **b** The resistance curves for Ti Sample 2. **c** The resistance curve measured at 248 GPa, where the derivative of the resistance with respect to temperature $dR/dT$ is plotted to clearly show the onset $T_c$.

**Fig. 2. The superconductivity at magnetic field. a** Temperature dependence of the electrical resistance of Ti metal measured under different magnetic fields at the fixed pressure of 248 GPa. **b** Upper critical field versus superconducting transition

temperature of $T_c^{zero}$. The line is a fit obtained using the Ginzburg–Landau function.

**Fig. 3. The Hall measurements. a** and **b** Hall resistance as a function of magnetic field measured under different pressure. **c** and **d** Carrier density and resistance at fixed temperature versus pressure, respectively.

**Fig. 4. Superconducting phase diagram.** The superconducting critical transition temperature ($T_c$) of Ti metal at compression up to 320 GPa versus high pressure phases.

**Fig. 5. Calculated electronic band structures. a** Ti-ω at 20 GPa, **b** Ti-γ at 100 GPa and **c** Ti-δ at 180 GPa. The contributions from the *s*-electron and *d*-electron states are shown by red and gray circles, respectively, and the circle areas are proportional to the weights of the corresponding band states. Energy is measured relative to the Fermi energy *E*.

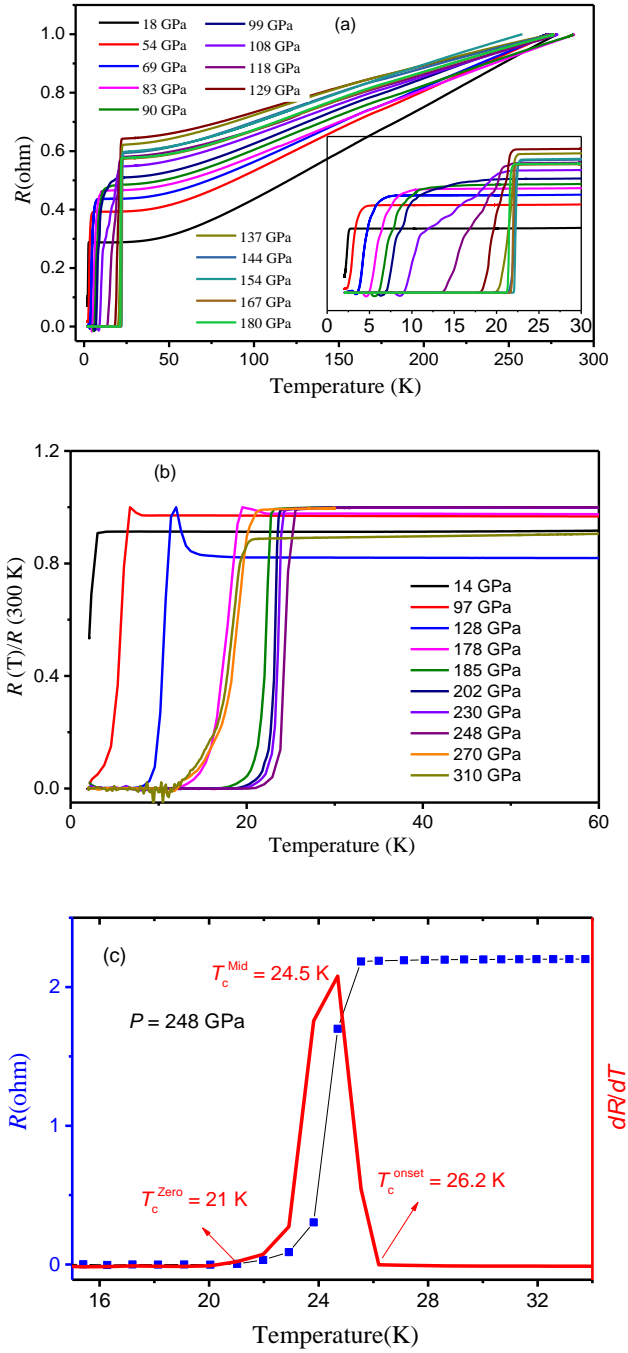

**Fig. 1. The superconductivity measurements. a** Temperature dependence of the electrical resistance of elemental metallic Ti (Sample 1) measured under high pressures. The inset is an enlarged view of the resistance curve, showing the superconducting transition in detail. **b** The resistance curves for Ti Sample 2. **c** The resistance curve measured at 248 GPa, where the derivative of the resistance with respect to temperature $dR/dT$ is plotted to clearly show the onset $T_c$.

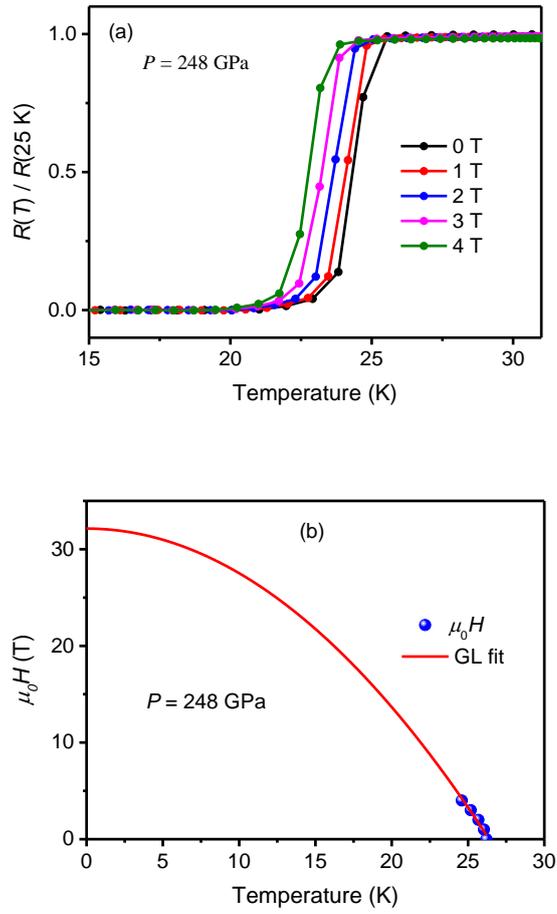

**Fig. 2. The superconductivity under magnetic field. a** Temperature dependence of the electrical resistance of Ti metal measured under different magnetic fields at the fixed pressure of 248 GPa. **b** Upper critical field versus superconducting transition temperature of $T_c^{zero}$. The line is a fit obtained using the Ginzburg-Landau (GL) function.

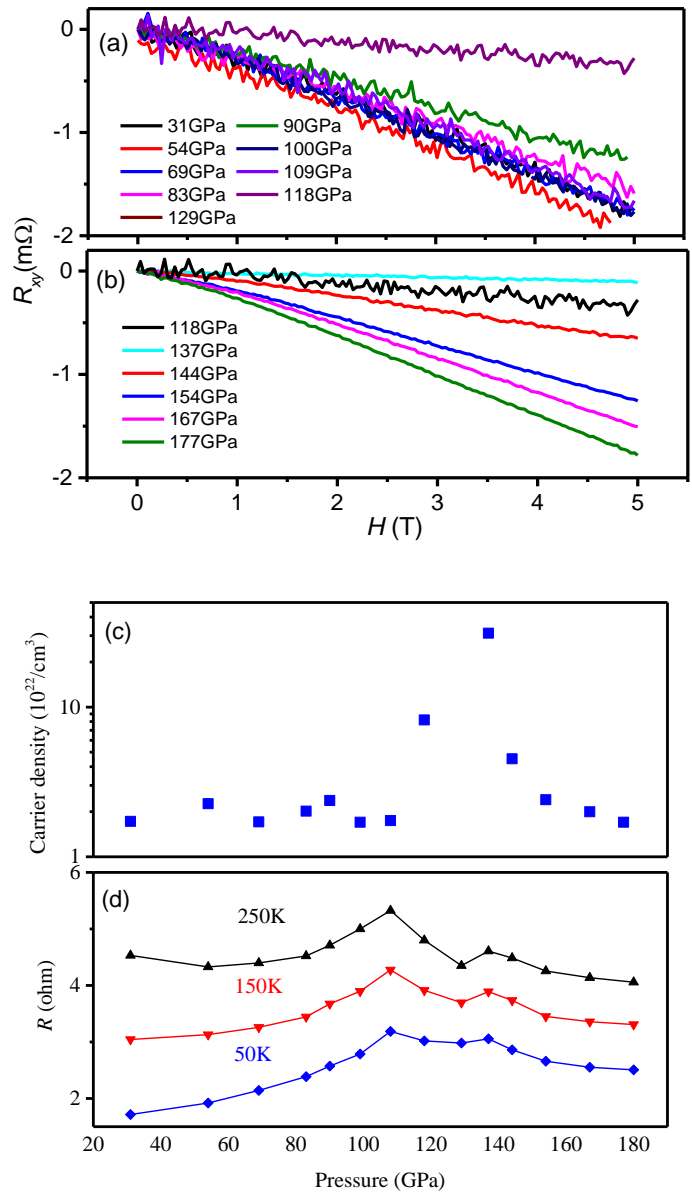

**Fig. 3. The Hall resistance measurements. a** and **b** Hall resistance as a function of magnetic field measured under different pressure. **c** and **d** Carrier density and resistance at fixed temperature versus pressure, respectively.

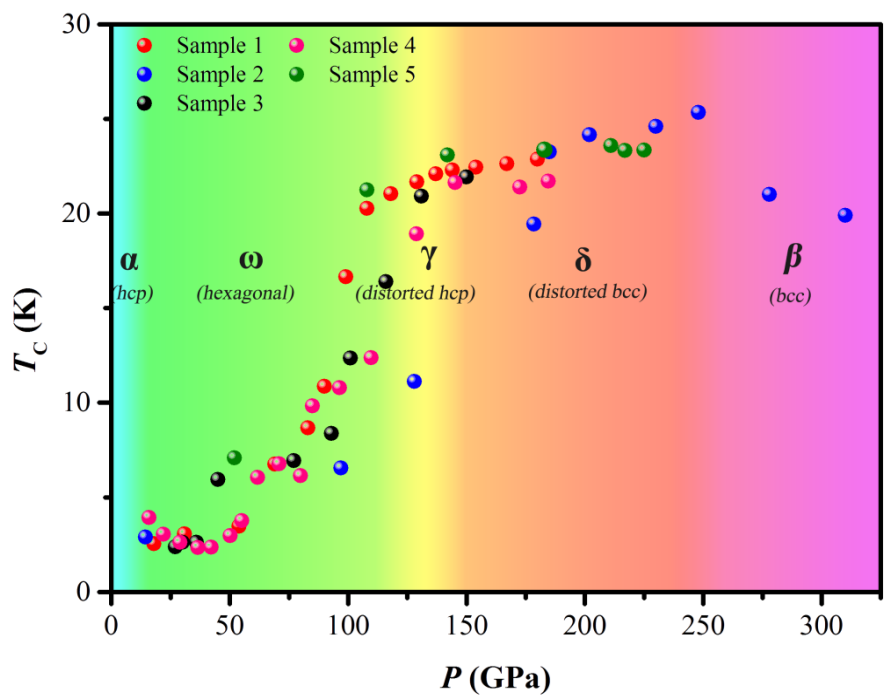

**Fig. 4. Superconducting phase diagram.** The superconducting critical transition temperature ($T_c$) of Ti metal under compression up to 320 GPa plotted against the background of stability fields of the indicated structural phases.

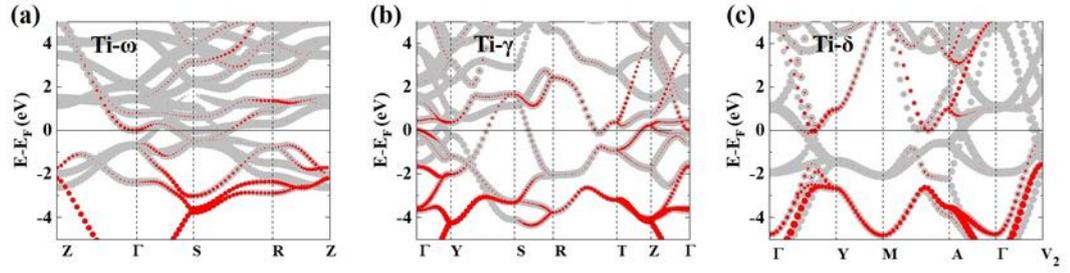

**Fig. 5. Calculated electronic band structures. a** Ti-ω at 20 GPa, **b** Ti-γ at 100 GPa and **c** Ti-δ at 180 GPa. The contributions from the *s*-electron and *d*-electron states are shown by red and gray circles, respectively, and the circle areas are proportional to the weights of the corresponding band states. Energy is measured relative to the Fermi energy $E_F$.

# Supplementary Materials

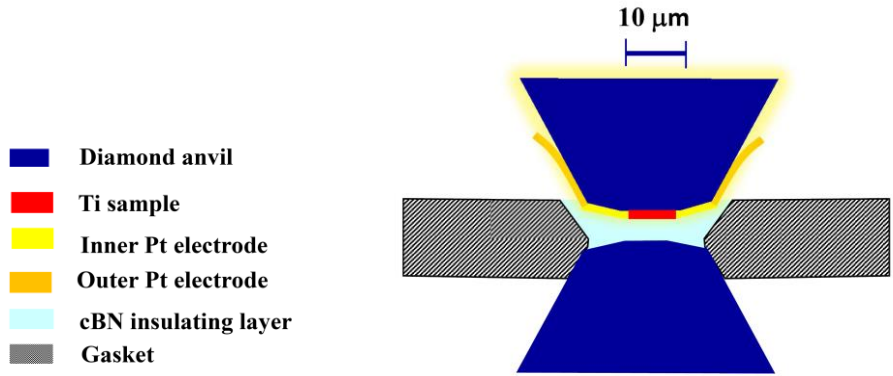

(a)

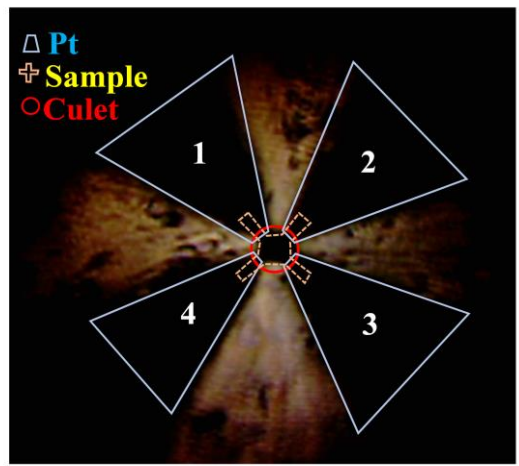

(b)

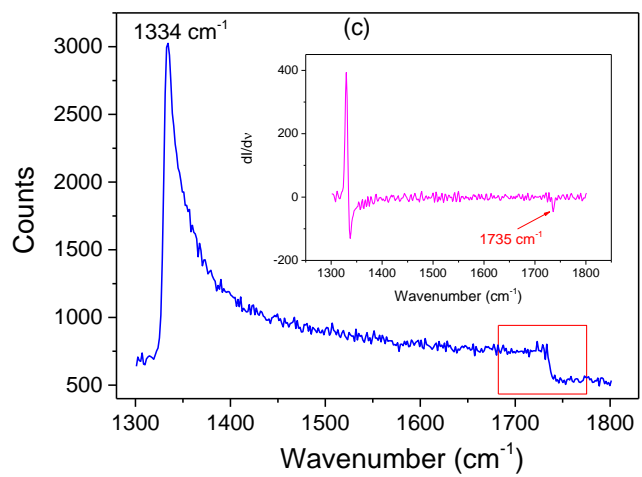

(c)

**Supplementary Figure 1.** (a) The schematic view of the high pressure assembly for diamond anvil cell. (b) The electrode configuration for four probes Van der Pauw method. (c) Typical Raman spectrum from the center of the diamond anvil culet at high pressure. The inset is the differential spectrum, clearly showing the Raman edge frequency.

The electrical resistance and Hall effect measurements were performed using the four-probe Van der Pauw method as shown in Supplementary Fig. 1a The standard way is to use 6 probes to measure Hall signals. However concerning the 30μm sample space as well as tiny specimen(<10 μm) in megabar pressure chamber four probes based on Van der Pauw method are sometimes applied. In our manuscript, the resistance and Hall coefficient are measured via the four probe Van der Pauw method, which is usually used for small samples with irregular and two-dimensional shape[L. van der Pauw J., "A Method of Measuring Specific Resistivity and Hall Effect of Discs of Arbitrary Shape", Philips Research Reports 13, 1(1958); Errandonea, D. *et al*. "Investigation of conduction-band structure, electron scattering mechanisms, and phase transitions in indium selenide by means of transport measurements under pressure", Physical Review B 55, 16217(1997); Zhang, J. K. *et al*. "Electronic topological transition and semiconductor to metal conversion of $Bi_2Te_3$ under high pressure", Applied Physics Letters 103, 52102(2013)]. For all the samples in the experiments, the electrode and the sample configurations are the same as shown in Supplementary Fig. 1b. For resistance experiments, the electrodes of 1 and 4 can be used for applying current, and the other electrodes of 2 and 3 for the longitudinal voltage measurement. While for Hall coefficient measurements, the electrodes of 1 and 3 are used for applying current, and the other electrodes of 2 and 4 for the Hall voltage measurement.

The pressures are calibrated via the shift of the first-order Raman edge frequency from the diamond cutlet by the following equation[18]. There are two first-order Raman edges at low and high frequencies of 1344 cm$^{-1}$ and 1735 cm$^{-1}$, which come from the table face (ambient pressure) and culet (high pressure) of the anvil, respectively.

$$P(\text{GPa}) \cong K_0 \frac{\Delta \nu}{\nu_0}\left[1 + \frac{1}{2}(K_0' - 1)\frac{\Delta \nu}{\nu_0}\right] \quad (1)$$

Here, $K_0$ = 547 GPa and $K_0$' = 3.75 are the bulk modulus and pressure derivative of bulk modulus of diamond, respectively. $\nu_0$ and $\nu$ are the edge frequencies at ambient pressure and high pressure, respectively. $\Delta \nu$ is the difference between $\nu_0$ and $\nu$.

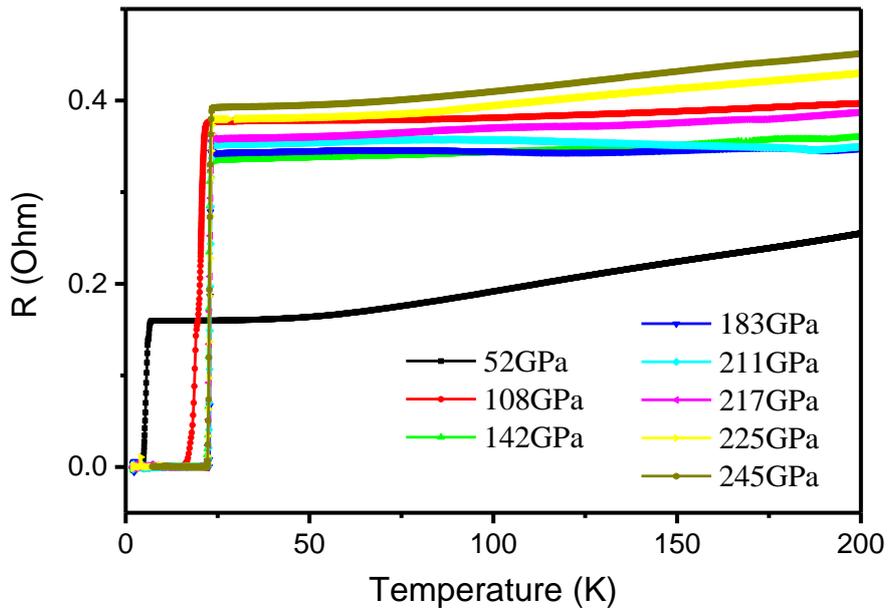

**Supplementary Figure 2. The superconductivity measurements.** Temperature dependence of the electrical resistance of elemental metallic Ti (Sample 5) measured under high pressures.

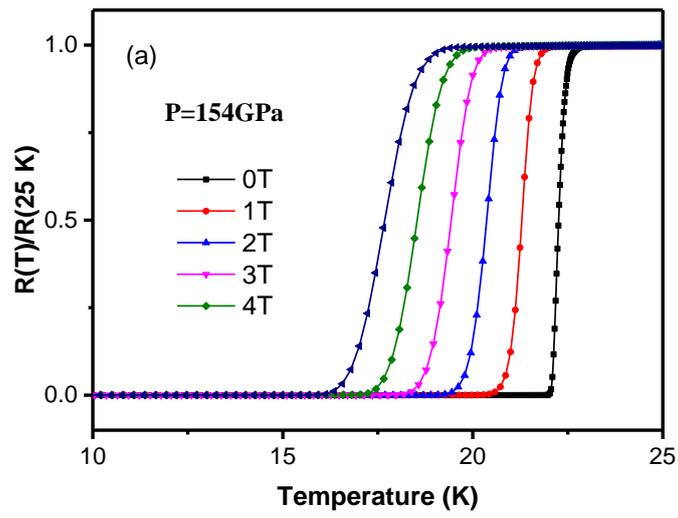

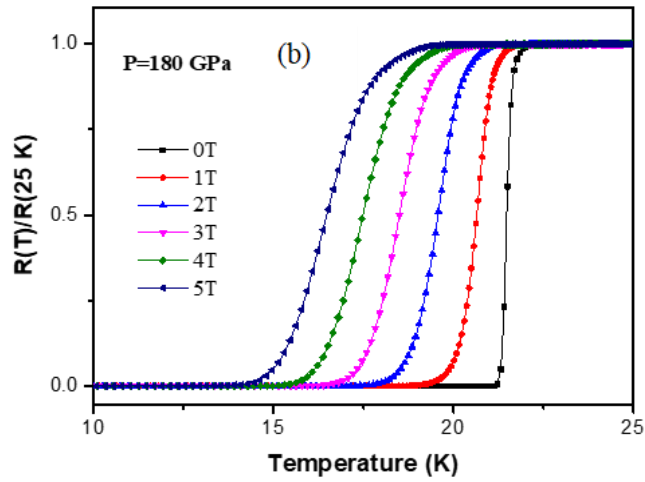

**Supplementary Figure 3. The superconductivity under magnetic field.** (a) The superconducting transitions measured under different magnetic fields at 154 GPa and (b) 180 GPa.

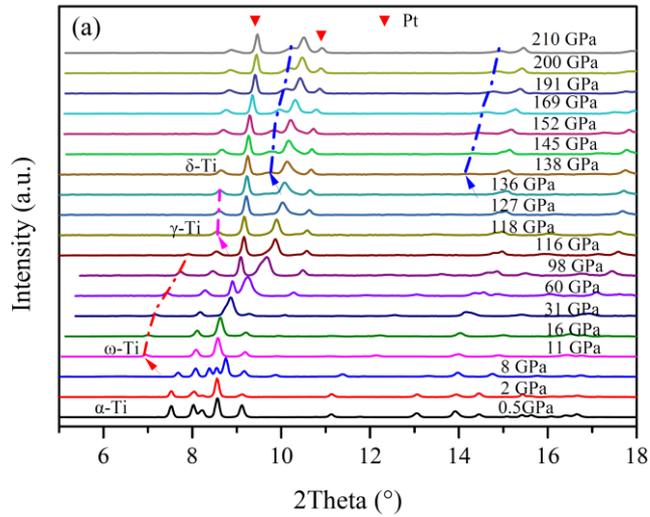

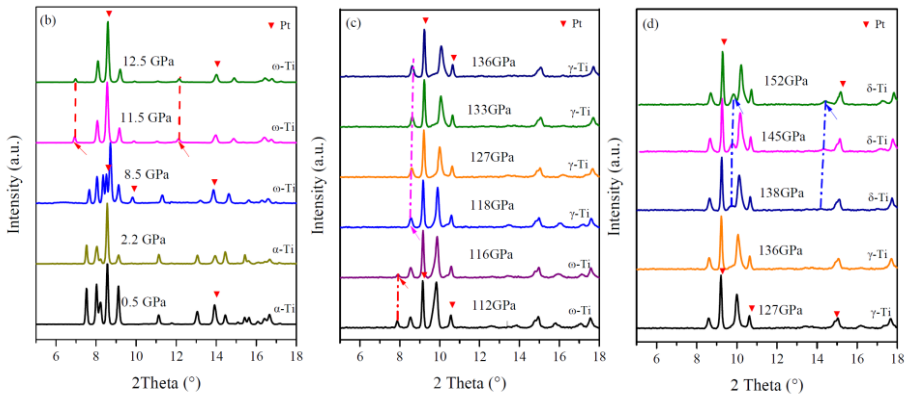

**Supplementary Figure 4. The X-ray diffraction and phase transition.** (a) The *in-situ* high pressure X-ray diffraction patterns with the highest experimental pressure of 210 GPa and the wavelength of 0.3344 Å. (b-d) The diffraction patterns with the pressure of 0.5-12.5 GPa, 112-136 GPa and 127-152 GPa, respectively. The Ti-α to Ti-ω phase transition at ~11.5 GPa can be evidenced with the appearance of two new peaks at 6.94 ° and 12.12 °. The disappearance of the peak at 7.90° above 116 GPa hints the Ti-ω to Ti-γ phase transition. The Ti-γ to Ti-δ phase transition occurs at 138 GPa, where two new peaks at ~9.72° and ~14.18 ° appear. The pressure induced phase transition sequence from Ti-α phase, to Ti-ω phase, to Ti-γ phase, and then to Ti-δ phase has been observed, which is consistent with the previously reported results

[6].

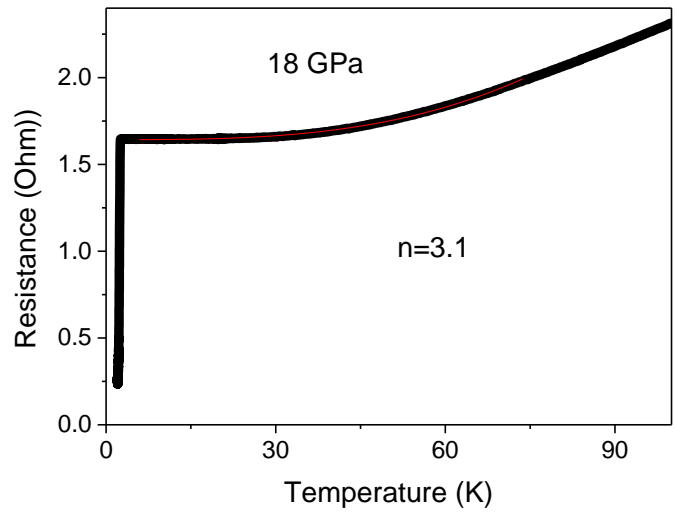

**Supplementary Figure 5. The *s-d* interband scattering.** The temperature dependence of resistance measured at 18 GPa. The resistance in the low temperature range of 6-70 K is fitted (red line) with the formula of $R= R_0+AT^n$, where $R_0$ is the residual resistance, $A$ is the coefficient of the power law, and $n$=3.1 is the extracted exponent. The *n* value near 3 implies that the *s-d* interband scattering dominates the electron transport with contributions from electron correlation effects.

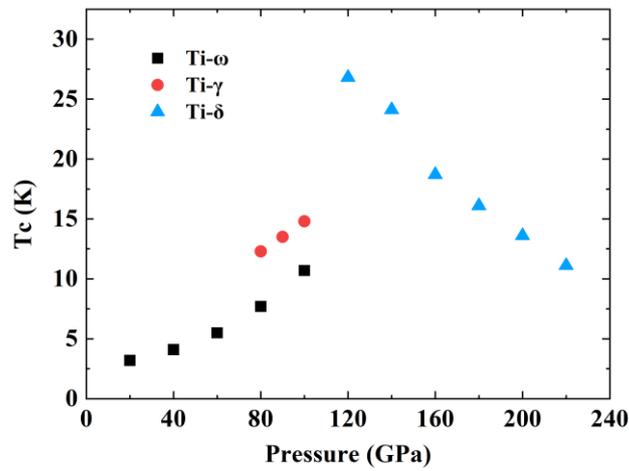

**Supplementary Figure 6. The calculated $T_c$ versus pressure.** The superconducting critical transition temperature ($T_c$) of the indicated Ti phases in their stability fields in the [20, 220] GPa hydrostatic pressure range. The calculated crystal structure of Ti is sensitive to the pressure conditions, and under hydrostatic pressures at or above 160 GPa, the Ti-δ phase essentially relaxes to the Ti-β (bcc) phase without considering any stress anisotropy.